\documentclass[aps,prl,twocolumn,showpacs,amsmath,amssymb,groupedaddress]{revtex4}
\usepackage{graphicx}
\usepackage{dcolumn}
\usepackage{bm}
\usepackage{amsfonts}
\newcommand{\ket}[1]{\mbox{$\mid \! #1 \, \rangle$}}

\begin{document}
\title{Experimental filtering of two-, four-, and six-photon singlets from
single PDC source}
\author{Magnus R\aa dmark$^{1}$,  Marcin Wie\'sniak$^{2}$, Marek \.{Zukowski}$^{2}$ and Mohamed
Bourennane$^{1}$}
\affiliation{ $^{1}$Physics Department, Stockholm University, SE-10691 Stockholm, Sweden\\
$^{2}$Institute for Theoretical Physics and Astrophysics, Uniwersytet
Gda\'{n}ski, PL-80-952 Gda\'{n}sk, Poland}
\date{\today}
\begin{abstract}
Invariant entangled states remain unchanged under simultaneous
identical unitary transformations of all their subsystems. We
experimentally generate and characterize such  invariant two-,
four-, and six-photon polarization entangled states. This is done
only with a suitable filtering procedure of multiple emissions of
entangled photon pairs from a single source, without any
interferometric overlaps. We get the desired states utilizing
bosonic emission enhancement due to indistinguishability. The setup
is very stable, and gives high interference contrasts. Thus, the
process is a very likely candidate for various photonic
demonstrations of  quantum information protocols.
\end{abstract}
\maketitle
%
Entanglement is an essential tool in many quantum information tasks.
Entangled  states of two qubits proved to be useful in various
quantum communication protocols like quantum teleportation, quantum
dense coding, and quantum cryptography. They are  the essence of
the first versions of Bell's theorem\cite{review}. However, the
expansion of quantum information science has now reached a state in
which many schemes are involving multiparty processes, and could
require multiqubit entanglement.

There is an interesting series of multiqubit states,  $\ket{\Psi_{k}^{-}}$,  where $k=2,4,6$ or more. They are invariant
under actions consisting of identical unitary
transformations of each  qubit \cite{ZR97a}:
$
U^{\otimes k}\ket{\Psi_{k}^{-}} = \ket{\Psi_{k}^{-}},
$
where $U^{\otimes k} = U\otimes...\otimes U$ denotes a tensor
product of $k$ identical unitary operators $U$. The property protects the states
against collective noise. The states are useful e.g. for communication
of quantum information between observers who do not share a
common reference frame \cite{BRS03}: any
realignment of the receiver's reference frame corresponds to an
application of the same transformation to each of the sent qubits.
The states  $\ket{\Psi_{k}^{-}}$ can also be used for secure quantum
multiparty cryptographic protocols, such as the  multi-party secret
sharing protocol \cite{HBB99,GKBW07}.

We generate correlations which characterize the six-photon
$\ket{\Psi_{6}^{-}}$ entangled state. This is done in a six photon interference experiment.
A six-photon interference was reported recently in \cite{LZGGZYGYP07}. To obtain graphs
states the authors of ref. \cite{LZGGZYGYP07} used {\em three } pulse-pumped
parametric down-conversion (PDC) crystals, and interferometric
overlaps, to entangle independently emitted pairs (each
from a different crystal) with each other. Schemes of this kind are generalizations of those of ref.  \cite{ZHWZ97}.
However, the overlaps make the scheme fragile.

In our experiment, by pulse pumping just {\em one crystal} and
extracting the right order process via suitable filtering and
beamsplitting (the method of \cite{ZZW95}), we observe
simultaneously effects attributable to the multi-photon invariant
entangled states $\ket{\Psi_{2}^{-}}$, $\ket{\Psi_{4}^{-}}$, and
$\ket{\Psi_{6}^{-}}$. The setup has no overlaps and therefore no
interferometric alignment is  needed. It is  strongly robust, and
the output is of high fidelity with respect to the theoretical
states $\ket{\Psi_{k}^{-}}$.

A simple quantum optical description of two phase matched modes, of the multiphoton state
that results out of a single pulse acting on a type-II PDC crystal,
can be put as
\begin{equation}
C \sum_{n=0}\frac{1}{n!}[-i\alpha(a_{0H}^{\dagger}b_{0V}^{\dagger} -
a_{0V}^{\dagger}b_{0H}^{\dagger})]^n\ket{0}. \label{emission}
\end{equation}
The symbol $a_{0H}^{\dagger}$ ($b_{0V}^{\dagger}$) denotes a creation
operator for one horizontal, $H$,  (vertical, $V$) photon in mode $a_{0}$
($b_{0}$), {\em etc.} $C$ is a normalization constant, the coupling parameter $\alpha$
is a function of pump power, non-linearity and length of the
crystal. This is a good approximation of the actual state, provided one
collects the photons under conditions that allow full
indistinguishability between separate two-photon emissions
\cite{ZZW95}. First, second, and  third order terms in the
 expansion in eq.~(\ref{emission}), correspond to an
emission of two, four, and  six photons, respectively, into two
spatial modes. These terms can be re-interpreted as the following
superpositions of photon number states:
\begin{align}
&\ket{1H_{a_{0}},1V_{b_{0}}}- \ket{1V_{a_{0}},1H_{b_{0}}},
\label{particle2}
\end{align}
\begin{align}
&\ket{2H_{a_{0}},2V_{b_{0}}}-
\ket{1H_{a_{0}},1V_{a_{0}},1V_{b_{0}},1H_{b_{0}}}
\nonumber \\
&+\ket{2V_{a_{0}},2H_{b_{0}}}, \label{particle4}
\end{align}
\begin{align}
&\ket{3H_{a_{0}},3V_{b_{0}}} -
\ket{2H_{a_{0}},1V_{a_{0}},2V_{b_{0}},1H_{b_{0}}}+
\nonumber \\
&+\ket{1H_{a_{0}},2V_{a_{0}},1V_{b_{0}},2H_{b_{0}}}-\ket{3V_{a_{0}},3H_{b_{0}}},
\label{particle3}
\end{align}
where e.g. $2H_{a_{0}}$ and $3H_{a_{0}}$ denotes two and three
horizontally polarized photons in mode $a_{0}$, respectively, {\em
etc.} The second and third order PDC is intrinsically different than
simple products of two and three entangled pairs. Due to the bosonic
nature of photons,  emissions of completely indistinguishable
photons are more likely,  than the ones giving birth to photons with
orthogonal polarization.

We report a {\em joint} observation, in one setup,  of the
correlations of the invariant two, four and six-photon polarization
entangled states given by the following superpositions:
$
\ket{\Psi_{2}^{-}} = \frac{1}{\sqrt{2}}(\ket{HV} -\ket{VH}),
$

\begin{equation}
\ket{\Psi_{4}^{-}} = \frac{2}{\sqrt{3}}\ket{GHZ_{4}^{+}} -
\frac{1}{\sqrt{3}}\ket{EPR}\ket{EPR}, \label{state4}
\end{equation}
and
\begin{equation}
\ket{\Psi_{6}^{-}} = \frac{1}{\sqrt{2}}\ket{GHZ_{6}^{-}} +
\frac{1}{2}(\ket{\overline{W}_{3}}\ket{W_{3}}
-\ket{W_{3}}\ket{\overline{W}_{3}}). \label{state6}
\end{equation}
The states in the superpositions are given by:
$$\ket{GHZ_{4}^{+}}=\frac{1}{\sqrt{2}}(\ket{HHVV}+\ket{VVHH})/\sqrt{2}$$, $$\ket{GHZ_{6}^{-}}=(\ket{HHHVVV}-\ket{VVVHHH}),$$ and
$$\ket{EPR}=\frac{1}{\sqrt{2}}(\ket{HV}+ \ket{VH}).$$
Finally
$$\ket{W_{3}}=\frac{1}{\sqrt{3}}(\ket{HHV}+\ket{HVH}+\ket{VHH}).$$
The ket $\ket{\overline{W}}$ is the spin-flipped  $\ket{W}$. The
states (\ref{state2}-\ref{state6}) are obtained out of different
orders of the PDC emission (figure 1), by selecting specific double,
quadruple and six-fold coincidences.

\begin{figure}
\includegraphics[width= \columnwidth]{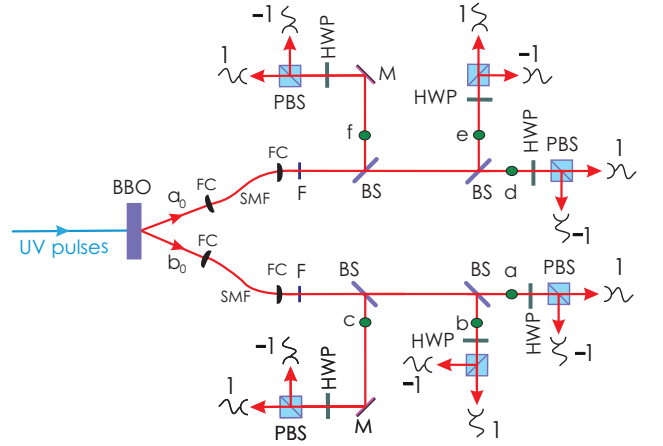}
\caption{\label{setup}Experimental setup for generating and
analyzing the six-photon polarization-entangled state.  The six
photons are created in third order PDC processes in a 2 mm thick BBO
pumped by UV pulses. The intersections of the two cones obtained in
non-collinear type-II PDC are coupled to single mode fibers (SMF)
wound in polarization controllers. Narrow band interference filters
(F) ($\Delta\lambda =3$ nm) serve to remove spectral
distinguishability. The coupled spatial modes are divided into three
modes each by 50\%-50\% beam splitters (BS). Each mode can be
polarization analyzed  using half wave plates
   (HWP) and a polarizing beam splitter (PBS). Simultaneous detection
   of six photons (two single photon detectors for each mode) are being recorded
   by a twelve channel coincidence counter.}
\end{figure}

In our setup we use a frequency-doubled Ti:Sapphire laser ($80$
Mhz repetition rate, $140$ fs pulse length) yielding UV pulses with
a central wavelength at $390$ nm and an average power of $1300$ mW.
The pump beam is focused to a $160$ $\mu$m waist in a $2$ mm thick
BBO ($\beta$-barium borate) crystal. Half wave plates and two $1$ mm
thick BBO crystals are used for compensation of longitudinal and
transversal walk-offs. The third order emission of non-collinear
type-II PDC is then coupled to single mode fibers (SMF), defining
the two spatial modes at the crossings of the two frequency
degenerated PDC emission cones. Leaving the fibers the
down-conversion light passes narrow band  ($\Delta\lambda =3$ nm)
interference filters (F) and is split into six spatial modes $(a, b,
c, d, e, f)$ by ordinary $50\%-50\%$ beam splitters (BS), followed
by birefringent optics (to compensate phase shifts in the BS's). Due
to the short pulses, narrow band filters, and single mode fibers the
down-converted photons are temporally, spectrally, and spatially
indistinguishable \cite{ZZW95}, see Fig.~\ref{setup}. The
polarization is being kept by passive fiber polarization
controllers. Polarization analysis is implemented by a half wave
plate (HWP), a quarter wave plate (QWP), and a polarizing beam
splitter (PBS) in each mode. The outputs of the PBS's are lead to
single photon silicon avalanche photo diodes (APD) through multi
mode fibers. The APD's electronic responses, following photo
detections, are being counted by a multi-channel coincidence counter
with a $3.3$ ns time window. The coincidence counter registers any
coincidence event between the  APD's as well as single detection
events.

The states $\ket{\Psi_{k}^{-}} (k=2,4,6)$ exhibit perfect two, four,
and six qubit correlations. The correlation function is defined as
an expectation value of the product of  local polarization ``Pauli''
observables. If one limits the measurement to the local observables
$\cos{\theta_{l}}\sigma_{z}^{(l)} +\sin{\theta_{l}}\sigma_{x}^{(l)}
$ (with eigenvectors $\sqrt{1/2}(\ket{L}_{l}\pm
e^{i\theta_{l}}\ket{R}_{l})$ and eigenvalues $\pm 1$),  the
measurements  correspond to linear polarization analysis in each
spatial mode ($l=a,b,c,d,e,f$). In such a case the quantum
prediction for the two photon (in modes b and d) correlation
function reads: $ E(\theta_{b},\theta_{d})= -\cos(\theta_{b} -
\theta_{d}). $ For the four photon counts (in modes a, b, d, and e)
correlation function is given by
\begin{align}
&E(\theta_{a},\theta_{b},\theta_{d},\theta_{e})=
\frac{2}{3} \cos(\theta_{a} + \theta_{b} - \theta_{d} - \theta_{e}) \nonumber \\
&+\frac{1}{3}\cos(\theta_{a} - \theta_{b})\cos(\theta_{c} -
\theta_{d}). \label{eq.correlation4}
\end{align}
Finally for the six photon events one has
\begin{align}
&E(\theta_{a},\theta_{b},\theta_{c},\theta_{d},\theta_{e},\theta_{f})= \nonumber \\
&-\frac{1}{2} \cos(\theta_{a} + \theta_{b} + \theta_{c} - \theta_{d}- \theta_{e} - \theta_{f} ) \nonumber \\
&-\frac{1}{18}\sum{\cos(\theta_{a} \pm \theta_{b} \pm \theta_{c} \pm
\theta_{d} \pm \theta_{e} \pm \theta_{f} )}, \label{eq.correlation6}
\end{align}
where $\sum$ is a sum over all possible sign sequences which contain
only {\em two positive} signs, with the sign sequence in the first term, proportional to $\frac{1}{2}$,
excluded. Due to the invariance, the
correlation functions for all measurements around any single great circle of the
Bloch sphere look the same.

\begin{figure}
\includegraphics[width= \columnwidth]{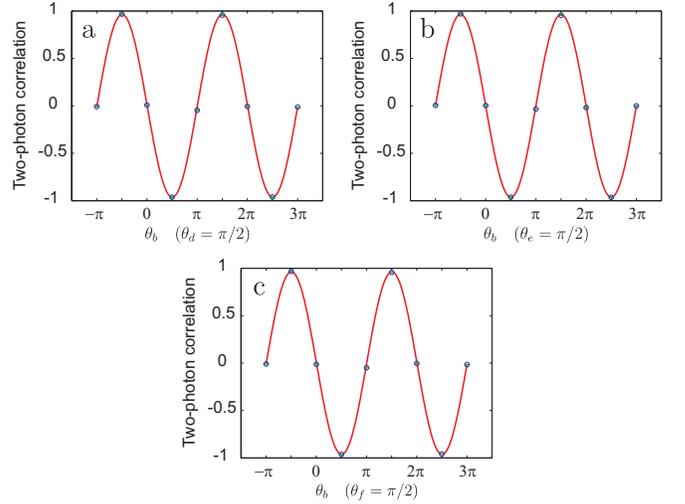}
\caption{\label{correlation2}Two-photon polarization correlation
 function. Modes $a, d, e$ and $f$ are analyzed in $D/A$ basis and mode $b$ analysis
 basis is varied around the equator of the Bloch sphere ($\sigma_{z} cos(\theta_{b})+\sigma_{x}
sin(\theta_{b})$). The solid lines show  sinusoidal fits to the
experimental data with a average visibility of $V_2 = 0.962\%\pm
0.003\%$.}
\end{figure}

Fig.~\ref{correlation2} shows  three experimentally observed
two-photon correlation functions, $E(\theta_{b},\theta_{l})$, where
$( l= d,e,f)$. The setting $\theta_{b}$ is varied,  while the other
analyzer is fixed at $\theta_{l}= \theta_{m}= \theta_{n}=\pi/2$.
This corresponds to diagonal/antidiagonal, $D/A$, linear
polarization analysis. A sinusoidal least-square fit was made to the
data. The average visibility, defined here, as the average amplitude
of the three fits,  is $V_2 = 0.962\%\pm 0.003\%$.

\begin{figure}
\includegraphics[width= \columnwidth]{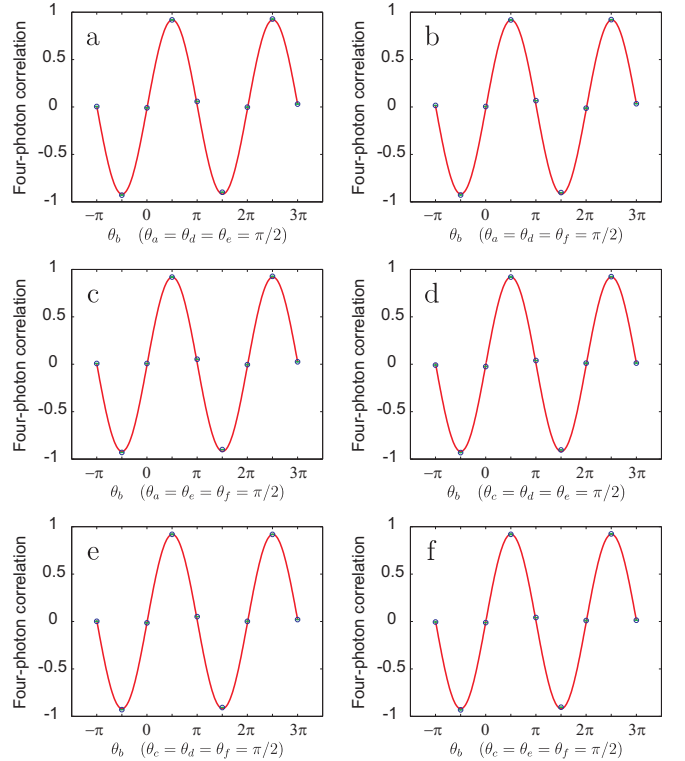}
\caption{\label{correlation4}Four-photon polarization correlation
functions of $\ket{\Psi_{4}^{-}}$. Modes $a, c, d, e$ and $f$ are
analyzed in the $+/-$-basis and mode $b$ analysis basis is rotated
around the equator of the Bloch sphere
($\hat\sigma_{z}\cos\theta_{b}+\hat\sigma_{x}\sin\theta_{b}$). The
figures correspond to different implementations of the state, using
different modes ($abde$, $abdf$, $abef$, $bcde$, $bcdf$ and $bcef$).
The solid lines show sinusoidal fits to the experimental data. The
solid lines show  sinusoidal fits to the experimental data with a
average visibility of $V_4 = 0.9189\%\pm0.0049\%$.}
\end{figure}

Fig.~\ref{correlation4}
shows how six experimentally observed four photon correlation
functions $E(\theta_{b},\theta_{l}, \theta_{m}, \theta_{n})$
 depend on $\theta_{b}$. The other
analyzers were fixed at $\theta_{l}=\pi/2,$ where $(l = a,c)$, $(m =
d,e)$, and $(n = e,f)$. The average value of the six visibilities is
$V_4 = 0.9189\%\pm0.0049\%$.

\begin{figure}
\includegraphics[width= \columnwidth]{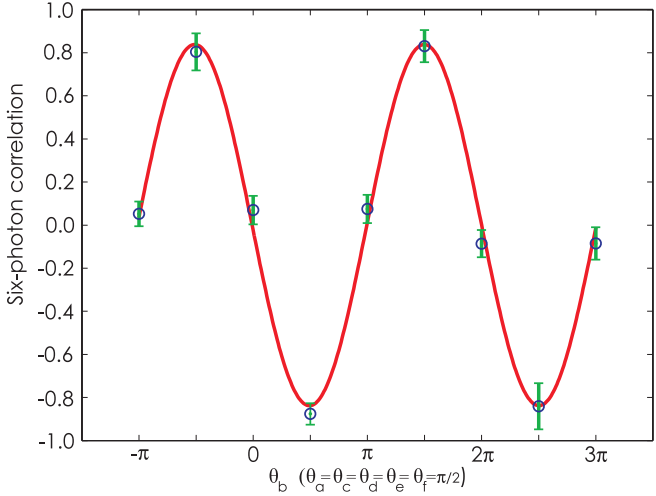}
\caption{\label{correlation6}Six-photon polarization correlation
 function. Modes $a, c, d, e$ and $f$ are analyzed in $D/A$ basis and
 mode $b$ analysis
 basis is varied around the equator of the Bloch sphere ($\sigma_{z} cos(\theta_{b})+\sigma_{x}
sin(\theta_{b})$). The solid line shows a sinusoidal fit to the
experimental data with a visibility of $83.79\%\pm 2.98\%$.}
\end{figure}

Finally, Fig.~\ref{correlation6} shows similar data for  the
experimentally observed six photon correlation function
$E(\theta_{b},\theta_{a}, \theta_{c},
\theta_{d},\theta_{e},\theta_{f})$. Again $\theta_{b}$ was varied
with the other five analyzers fixed at $\theta_{l}= \pi/2$ where $(l
= a,c,d,e,f)$. The value on the visibility is $V=83.79\%\pm2.98\%$.

In table \ref{t1} we present all the experimentally obtained two-,
four-, and six- photon visibilities.

\begin{table}
\caption{\label{tab:visibility}Visibilities of the invariant states
$\ket{\Psi_k^{-}}$ (where $k = 2, 4, 6$)}
\begin{ruledtabular}
\begin{tabular}{ccc}
$k$ & Modes  & Visibility  \\ \hline
2 & $b,d$ & $0.962\pm0.004$ \\
2 & $b,e$ & $0.963\pm0.006$ \\
2 & $b,f$ & $0.962\pm0.004$ \\
4 & $a,b,d,e$ & $0.919\pm0.014$ \\
4 & $a,b,d,f$ & $0.918\pm0.011$ \\
4 & $a,b,e,f$ & $0.919\pm0.014$ \\
4 & $a,b,e,f$ & $0.919\pm0.014$ \\
4 & $b,c,d,f$ & $0.918\pm0.009$ \\
4 & $b,c,e,f$ & $0.920\pm0.012$ \\
6 & $a,b,c,d,e,f$ & $0.838\pm0.030$
\end{tabular}
\end{ruledtabular}
\label{t1}
\end{table}


We have compared the observed visibilities with theoretical
predictions, \cite{ARXIV}. To estimate maximal predictable
visibilities one can use a less  simplified description of the
two-photon state emitted by SPDC event, and replace in eq.
(\ref{emission}) $(a_{0H}^{\dagger}b_{0V}^{\dagger} -
a_{0V}^{\dagger}b_{0H}^{\dagger})$ by
\begin{eqnarray}
&& \int dt \int d\omega_0\int d\omega_1\int
d\omega_2f(\omega_1)f(\omega_2) g(\omega_0)e^{i\omega t}\times \nonumber\\
&& \Delta(\omega_0-\omega_1-\omega_2)
(a_{0H}^{\dagger}(\omega_1)b_{0V}^{\dagger}(\omega_2) -
a_{0V}^{\dagger}(\omega_1)b_{0H}^{\dagger}(\omega_2)). \nonumber\\ &&
\end{eqnarray}
This approach is rich enough to take into account the frequency
phase matching conditions. The creation operators depend
additionally on frequencies, and obey
$[a_{0X}(\omega),a_{0X'}(\omega')] =
\delta_{XX'}\delta_{XX'}(\omega- \omega')$, etc.
The function $f(\omega)$
represents the shape of the filter transmission profiles, and
$g(\omega_0)$ represents the frequency profile of the pump pulse. If
one chooses $f(\omega) = \exp[-((\omega_f -\omega)/(2\sigma_f))^2]$
and $g(\omega) = \exp[-((\omega_p -\omega)/(2\sigma_p))^2]$ where
$\sigma_p$, and $\sigma_f$ are the FWHM bandwidth of the pump, and
the filters, and $\omega_f = \omega_p/2$, the following formulas for
the maximal theoretical visibility as a function of the ratio $r =
\sigma_f/\sigma_p$ can be reached, \cite{ARXIV}. For the
four-photon process: $V^{temp}_4 = \sqrt{1+2r^2}/(1+r^2)$ and for
the six-photon process: $V^{temp}_6 = (1+2r^2)/[(1+r^2/2)(1+3r^2/2)]$.
In our experiment  we used $r_{exp} =
\Delta\lambda_f/(4\Delta\lambda_p) = 0.76$. This corresponds  to
$V_4 = 0.93$ and $V_6 = 0.90$. The actual measured values of
visibility for two, four, and six photon interference are very close
to the predicted ones, see table \ref{t1}.  Thus,  the fact that our setup use only
filtering and beamsplitting, has interferometric
advantages. In other words the obtained four  and six particles
visibilities are almost as high as one get for the ratio $r_{exp}$.

The high visibility has the following consequences. With just a part
of our
 data one can use the simplest one of the  ``experimentally friendly''
entanglement indicators introduced in \cite{BBLPZ08}. It
guarantees that N-qubit state is entangled, if the norm of the
N-particle correlation tensor is higher than 1. For the
$\ket{\Psi_{2}^{-}}$ we take just $T_{xx}$, $T_{yy}$, $T_{zz}$, for
$\ket{\Psi_{4}^{-}}$ again just $T_{xxxx}$, $T_{yyyy}$, $T_{zzzz}$,
and for $\ket{\Psi_{6}^{-}}$ the components $T_{xxxxxx}$,
$T_{yyyyyy}$ and $T_{zzzzzz}$. With our  data we have obtained the
 $2.785\pm 0.007$, $2.517\pm0.011$, and $2.29 \pm
0.14$, respectively for each of the case. The entanglement threshold
is violated by of $242$, $133$, and $9.3$ standard deviations.
Additionally,  according to the criteria given in \cite{ZB02} the
state cannot be described by a local realistic model, if the sum of
squares of two out the the listed components is above 1. This is
again achieved by our data: for four particles we get
$T_{xxxx}^2+T_{yyyy}^2=1.646\pm 0.009$ (exceeding $1$ by $74.8$
standard deviations)
  and for six particles we have $T_{xxxxxx}^2+T_{yyyyyy}^2=1.52\pm 0.11$
 (exceeding $1$ by $4,5$ standard deviations). Thus the state can be directly
 (that is without still enhancing
its fidelity) utilized in classical threshold beating communication
complexity protocols \cite{BZPZ04}.

In
summary, we have  experimentally demonstrated that  a suitable
filtering procedure applied to  a triple emission from a {\em
single} pulsed source of polarization entangled photons leads to two, four and
six-photon, high visibility, interference due to entanglement, observable in a single setup. We
utilize the bosonic emission enhancement occurring in the emission
of three photon-pairs in PDC, thus the
process is not entirely spontaneous. The six qubit
state that we observed is invariant with respect to
simultaneous identical (unitary) transformations of all qubits. This
makes it particularly useful for multiparty quantum communication and
general quantum computation tasks: it is very robust against
deformations in transfer. Since we use just one source, we avoid
alignment problems, and thus the setup is very stable. We would like
to note that the interference contrast is high enough for our source
to be used in two-, four-, and six party demonstrations of quantum
reduction of communication complexity in some joint computational
tasks, and for secret sharing, as well as in many other quantum
informational technologies. \\

This work was supported by
Swedish Research Council (VR).  M.\.{Z}. was supported by Wenner-Gren Foundations
and by the EU programme QAP (Qubit Applications, No. 015858). \\

\end{document}